\documentclass[twocolumn,aps,a4paper]{revtex4}
\usepackage{graphicx}

\begin{document}

\title{Uncertain growth and the value of the future}
\author{Jaume Masoliver}
\affiliation{Departament de F\'isica Fonamental, Universitat de Barcelona, Barcelona, Spain}, 
\author{Miquel Montero}
\affiliation{Departament de F\'isica Fonamental, Universitat de Barcelona,  Barcelona, Spain}, 
\author{Josep Perell\'o}
\affiliation{Departament de F\'isica Fonamental, Universitat de Barcelona,  Barcelona, Spain}, 
\author{John Geanakoplos}
\affiliation{Department of Economics, Yale University, New Haven CT USA}, 
\author{J. Doyne Farmer}
\affiliation{Mathematical Institute and Institute for New Economic Thinking at the Oxford Martin School, \\
University of Oxford, Oxford, UK}

\date{\today}

\begin{abstract}

For environmental problems such as global warming future costs must be balanced against present costs. This is traditionally done using an exponential function with a constant discount rate, which reduces the present value of future costs. The result is highly sensitive to the choice of discount rate and has generated a major controversy as to the urgency for immediate action.  We study analytically several standard interest rate models from finance and compare their properties to empirical data. From historical time series for nominal interest rates and inflation covering 14 countries over hundreds of years, we find that extended periods of negative real interest rates are common, occurring in many epochs in all countries. This leads us to choose the Ornstein-Uhlenbeck model, in which real short run interest rates fluctuate stochastically and can become negative, even if they revert to a positive mean value. We solve the model in closed form and prove that the long-run discount rate is always less than the mean; indeed it can be zero or even negative, despite the fact that the mean short term interest rate is positive. We fit the parameters of the model to the data, and find that nine of the countries have positive long run discount rates while five have negative long-run discount rates. Even if one rejects the countries where hyperinflation has occurred, our results support the low discounting rate used in the Stern report over higher rates advocated by others.

\end{abstract}

\maketitle

\section{Overview}

In economics ``discounting" refers to weighting the future relative to the present \cite{Samuelson}. 
The choice of a discounting function has enormous consequences for long run environmental
planning \cite{DasGupta2004}. For example, in a highly influential report on climate change commissioned by the UK government, Stern \cite{Stern} uses a discounting rate of $1.4\%$, which on a 100 year horizon implies a present value of $25\%$ (meaning the future is worth $25\%$ as much as the present). In contrast, Nordhaus \cite{Nordhaus} argues for a discount rate of $4\%$, which implies a present value of $2\%$, and at other times \cite{Nordhaus2007} has advocated rates as high as $6\%$, which implies a present value of $0.3\%$. The choice of discount rate is perhaps the biggest factor influencing the debate on the urgency of the response to global warming. Stern has been widely criticized for using such a low rate \cite{Nordhaus,Nordhaus2007,DasGupta2006,Mendelsohn,Weitzman2007,Nordhaus2008}. This issue is likely to surface again with the upcoming Calderon report in July 2014.

A simple argument to motivate discounting is based on opportunity cost. Under a constant, continuously compounded rate of interest $r$, a dollar invested today will yield $e^{rt}$ at time $t$, so an environmental problem that costs $X$ to fix at time $t$ is equivalent to an investment of $e^{-rt}X$ now. Economists present a variety of reasons for discounting, including impatience, economic growth, and declining marginal utility; these are embedded in the Ramsey formula, which forms the basis for the standard approaches to discounting \cite{ArrowReview}. Here we adopt the net present value approach, which treats the real interest rate as the measure of the trade-off between consumption today and consumption next year, without delving into the factors influencing the real interest rate. We estimate the stochastic real interest rate process using historical data \cite{ArrowReview,NewellPizer}. 

It is often argued that, based on past trends in economic growth, future technologies will be so powerful compared with present technologies that it is more cost-effective to
encourage economic growth, or to solve other problems such as AIDS or malaria, than it is to take action against global warming now \cite{Nordhaus2008}. Analyses supporting this conclusion typically study discounting by working with an interest rate that is fixed over time, ignoring fluctuations about the average. This is mathematically convenient, but it is also dangerous: In this problem, as in many others, fluctuations play a decisive role.

A proper analysis takes fluctuations in the real interest rate, caused partly by fluctuations in growth, into account. When the real interest rate $r(t)$ varies the discounting function becomes
\begin{equation}
D(t) = E \left[ \exp\left( {-\int_{0}^{t} r(t^{\prime})dt^{\prime}}\right)\right],
\label{D}
\end{equation}
where the expectation $E[\cdot]$ is an average over all possible interest rate paths. The fact that this is an average of exponentials, and not an exponential of an average, implies that the paths with the lowest interest rates dominate, and in general lowers $D(t)$. This has been shown in several ways. Early papers analyzed an extreme case in which the annual real rate is unknown today, but starting tomorrow will be fixed forever at one of a finite number of values \cite{Weitzman98,Gollier}. More recent papers simulate stochastic interest rate processes out to some horizon, leaving aside the asymptotic behavior of real rates \cite{NewellPizer,Groom,Hepburn,Freeman}. 

The presence of fluctuations can dramatically alter the functional form of the discounting function. If interest rates follow a geometric random walk, for example, the discounting function asymptotically decays as a power law of the form $D(t) = At^{-1/2}$ \cite{Farmer}. In contrast to the exponential function, this is not integrable on $(0, \infty)$, underscoring how important the effect of persistent fluctuations can be.

\section{Results}

To understand how discounting depends on the random process used to characterize interest rates, we have studied three different models which appear ubiquitously in the literature \cite{jouini} using both analytical and numerical methods. The models are: the Ornstein-Uhlenbeck (OU) process \cite{OU}, the Feller process \cite{feller} and the log-normal \cite{Osborne} process. In two of them (Feller and log-normal) rates cannot take negative values while in the OU model $r(t)$ can be either positive or negative. The analytical results which will be presented elsewhere are summarized in Table \ref{tab1} where we see that the discounting behavior depends sensitively on the choice of model and parameters\footnote{Sensitivity to parameters was also observed by Groom et al for a different set of models \cite{Groom}.}.

\begin{table*}
\begin{tabular}{lccl}
\hline
{\bf Models} & {\bf Rates} & {\bf Mean Reversion} & {\bf Long-run Discount Function} \\ \hline
 	& positive	& & (1) exponentially decreasing\\
Ornstein-Uhlenbeck & and & yes & (2) saturation to a constant value \\
	& negative & & (3) exponentially increasing \\\hline 	
Feller & positive & yes & (1) exponentially decreasing \\
	& & & (2) saturation to a constant value \\ \hline
	& 	& & (1) exponentially decreasing\\
Log-normal & positive & no & (2) power-law decreasing\\
	& & & (3) saturation to a constant value \\ \hline
\end{tabular}
\caption{Summary of the main properties and the long-time behavior of the discounting function $D(t)$ for three models of interest rates. Note that (i) mean reversion means the existence of a force drifting rates towards their average value.; (ii) the type of asymptotic regime depends on the values of the parameters appearing in each model; (iii) The O-U model allows for positive or negative interest rates, whereas the other two assume they are positive.}.
\label{tab1}
\end{table*}

\begin{figure}
\vspace*{.05in}
\centering
\includegraphics[width=8.9cm]{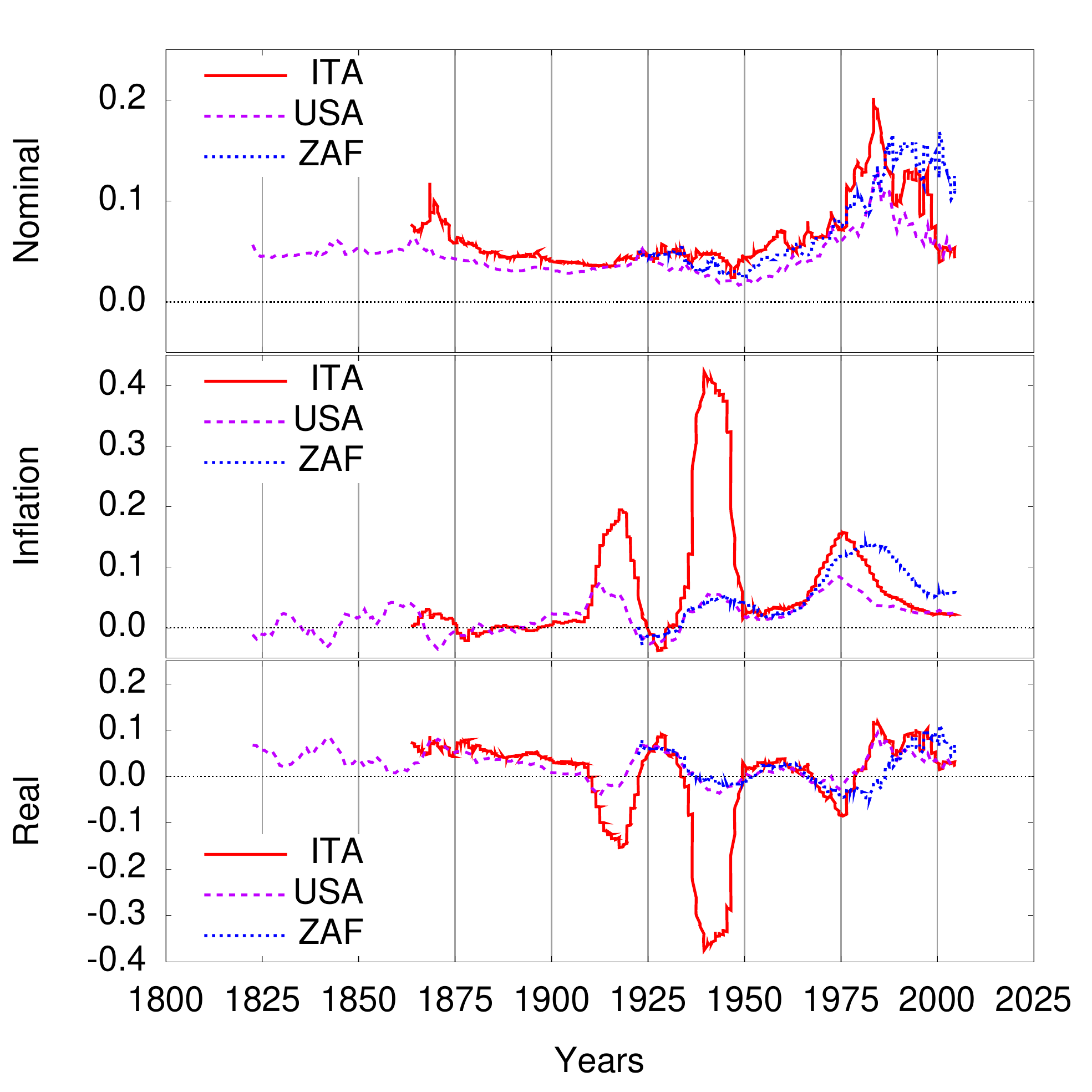}
\caption{Real interest rates display large fluctuations and negative rates are not uncommon. We show nominal interest rates (top), inflation (middle), and real interest rates (bottom) for 
Italy (ITA), United States (USA) and South Africa (ZAF). \label{timeSeries}}
\end{figure}

To determine which model is most appropriate we collected data for nominal interest rates and inflation for fourteen countries over spans of time ranging from 87 to 318 years and used these to construct real interest rates. The countries in our sample are: Argentina (ARG, 1864-1960), Australia (AUS, 1861-2012), Chile (CHL, 1925-2012), Germany (DEU, 1820-2012), Denmark (DNK, 1821- 2012), Spain (ESP, 1821-2012), United Kingdom (GBR, 1694-2012), Italy (ITA, 1861-2012), Japan (JPN, 1921-2012), Netherlands (NLD, 1813-2012), Sweden (SWE, 1868-2012), the United States (USA, 1820-2012), and South Africa (ZAF, 1920-2012). Some examples are plotted in Figure~\ref{timeSeries}. Since all but two of our nominal interest rate processes are for ten year government bonds, which pay out over a ten year period, we smooth inflation rates with a ten year moving average, and subtract the annualized inflation index from the annualized nominal rate to compute the real interest rate as described in the Appendix.

A striking feature observed in many epochs for all countries is that real interest rates frequently become negative, often by substantial amounts and for long periods of time (see Table \ref{tab3}). This immediately rules out most standard financial models, which assume that interest rates are essentially always positive. It also illustrates a central problem in previous work; all of the papers cited earlier assume real interest rates are always positive. We thus focus our attention on the Ornstein-Uhlenbeck model, which is the only one of the three models we have studied that allows negative interest rates. It can be written as
\begin{equation}
dr(t)=-\alpha(r(t)-m)dt+kdw(t),
\label{dr}
\end{equation}
where $r(t)$ is the real interest rate and $w(t)$ is a Wiener process. The parameter $m$ is a mean value to which the process reverts, $k$ is the amplitude of fluctuations, and $\alpha$ is the strength of the reversion to the mean.

Using Fourier-transform methods, in the Appendix we derive an exact solution for the discount function $D(t)$ of the time-dependent OU model. Letting $r_0=r(0)$ be the initial return, the probability density function $p(r, t | r_0)$  is a normal distribution, which in the large time limit has mean $m$ and variance 
\begin{equation}
\sigma^2 = k^2/2\alpha.
\label{stdv}
\end{equation}
In the limit $t \to \infty$ the discount function decays exponentially, i.e.
\begin{equation}
D(t)\simeq e^{-r_\infty t},
\label{assymptotic_D}
\end{equation}
 where
\begin{equation}
r_\infty = m-k^{2}/2\alpha^{2}.
\label{r_inf}
\end{equation}

\begin{table*}

\begin{tabular}{l l r r r r r r r r r r r r r}
\hline
Country & Neg RI & $m^{(-)}$\% & $m$\% &$1/\alpha$ & $k$x & $\mu$ & Min & Max & $\kappa$ & Min & Max & $r_\infty \%$ & Min & Max \\ \hline
Italy & $28\% \, (40y)$ & $13.3$ & $-0.3$ & $4.5$ & $6.9$ & $-0.01$ & $-0.42$ & $0.26$ & $0.68$ & 0.08 & 1.0 & $-5.4$ & $-20 $ & $5.5$ \\
Chile & $56\% \, (43y)$ & $25.1$ & $-6.8$ & $2.5$ & $25$ & $-0.17$ & $-0.50$ & $0.30 $ & $0.98$ & $0.22$ & $1.7$ & $-26$ & $-74$ & $10$\\
Canada & $22\% \, (20y)$ & $1.2$ & $2.9$ & $3.8$ & $2.3$ & $0.11$ & $0.00$ & $0.23$ & $0.18$ & $0.08$ & $0.15$ & $2.5$ & $0.0$ & $5.8$\\
Germany & $14\% \, (25y)$ & $100$ & $-10.7$ & $5.0$ & $34$ & $-0.55$ & $-2.6$ & $0.20$ & $3.9$ & $0.10$ & $7.1$ & $-160$ & $-540$ & $3.9$\\
Spain & $25\%\, (45y)$ & $3.0$ & $5.7$ & $17$ & $2.9$ & $0.96$ & $-0.08$ & $2.3$ & $2.0$ & $0.85$ & $2.5$ & $-6.4$ & $-4.8$ & $4.5$\\
Argentina & $20\%\, (17y)$ & $8.8$ & $2.4$ & $2.6$ & $6.2$ & $0.06$ & $-0.07$ & $0.18$ & $0.26$ & $0.11$ & $0.28$ & $1.1$ & $-4.4$ & $6.5$ \\
Netherlands & $17\% \, (33y)$ & $1.9$ & $3.2$ & $7.1$ & $1.6$ & $0.23$ & $0.06$ & $0.40$ & $0.34$ & $0.17$ & $0.44$ & $2.4$ & $-0.4$ & $5.0$\\
Japan & $33\% \, (26y)$ & $16.1$ & $-2.2$ & $4.2$ & $9.7$ & $-0.09$ & $-0.32$ & $0.17$ & $0.81$ & $0.09$ & $1.1$ & $-10$ & $-23$ & $3.9$ \\
Australia & $23\%\, (33y)$ & $2.7$ & $2.6$ & $5.3$ & $2.3$ & $0.14$ & $-0.04$ & $0.25$ & $0.27$ & $0.08$ & $0.33$ & $1.9$ & $-1.1$ & $4.8$\\
Denmark & $18\%\, (33y)$ & $1.7$ & $3.2$ & $4.3$ & $2.3$ & $0.14$ & $0.07$ & $0.18$ & $0.21$ & $0.10$ & $0.26$ & $2.7$ & $1.0$ & $4.0$\\
South Africa & $43\% \, (36y)$ & $0.6$ & $1.8$ & $4.8$ & $2.5$ & $0.08$ & $-0.10$ & $0.26$ & $0.26$ & $0.12$ & $0.21$ & $1.1$ & $-2.3$ & $5.1$\\
Sweden & $28\% \, (38y)$ & $1.9$ & $2.3$ & $4.0$ & $2.5$ & $0.09$ & $-0.01$ & $0.15$ & $0.20$ & $0.05$ & $0.27$ & $1.9$ & $-0.3$ & $3.8$\\
U.K. & $14\%\, (45y)$ & $0.1$ & $3.3$ & $5.3$ & $1.9$ & $0.18$ & $0.07$ & $0.23$ & $0.23$ & $0.12$ & $0.29$ & $2.8$ & $0.6$ & $4.0$ \\
U.S.A & $19\%\, (37y)$ & $1.8$ & $2.6$ & $5.6$ & $1.8$ & $0.14$ & $0.05$ & $0.22$ & $0.23$ & $0.16$ & $0.27$ & $2.1$ & $0.3$ & $3.8$\\\hline
All countries & $26\%\,(34y)$ & $12.8$ & $0.71$ & $5.4$ & $7.3$ & $0.09$ & $-0.28$ & $0.38$ & $0.75$ & $0.17$ & $1.14$ & $-13.6$ & $-48$ & $5.0$ \\
Stable coun.
& $23\%\,(33y)$ & $2.3$ & $2.7$ & $4.7$ & $2.6$ & $0.13$ & $0.00$ & $0.23$ & $0.24$ & $0.11$ & $0.28$ & $2.1$ & $-0.7$ & $4.8$ \\
Unstable coun. & $31\%\,(36y)$ & $32$ & $-2.9$ & $6.6$ & $16$ & $0.03$ & $-0.78$ & $0.65$ & $1.67$ & $0.27$ & $2.68$ & $-42$ & $-132$ & $5.6$ \\
\hline
\end{tabular}
\caption{A summary of our results showing how real interest rates result in a low long-run rate of discounting. This is driven by the fact that average real interest rate $m$ is typically low and the volatility $k$ is substantial. The fact that the characteristic time $1/\alpha$ is typically only a few years implies the long-run discounting rate $r_\infty$ is obtained quickly. Stable countries refer to those with positive $r_\infty$ and unstable countries to those with negative $r_\infty$. {\it Notes}: (i) ``Neg RI" gives the percentage of time and the total number of years in which real interest rates are negative. (ii) $m^{(-)}$ is the average amplitude (in percentage) during negative years only. (iii) $m$ is the mean real interest rate. (iv)
$1/\alpha$ is the characteristic reversion time in years. (v) $\kappa$ is the non-dimensional noise intensity in percent. (vi) $\mu$ is the non-dimensional mean interest rate. (vii) The Min and Max columns present the minimum and maximum by dividing each series into four equal blocks and estimating parameters separately for each block. (viii) $\kappa$ is the non-dimensional fluctuation amplitude. 
(ix) $r_\infty$ is the long-run real interest rate. Negative values of $r_\infty$ mean the discount function is asymptotically increasing.}
\label{tab3}
\end{table*}

Thus the long-run interest rate $r_\infty$ is always lower than the average interest rate $m$, by an amount that depends on the noise parameter $k$ and the reversion parameter $\alpha$. From equations (\ref{stdv}) and (\ref{r_inf}) it is evident that for any given mean interest rate $m$, by varying $k$ and $\alpha$ the long-run discount rate $r_\infty$ can take any desired value, including negative values. Furthermore, holding $m$ and $r_\infty$ fixed, the standard deviation $\sigma$ can be made arbitrarily small. This implies that the probability that $r(t) < r_\infty$ can be made arbitrarily small, even when $r_\infty \ll m$ (see Appendix). Note that the long run distribution $(m, \sigma)$ does not determine $r_\infty$ by itself; on the contrary, any $r_\infty < m$ is consistent with it. By increasing the persistence parameter $1/\alpha$ while holding the long run distribution $(m,\sigma)$ constant it is possible to lower $r_\infty$ to any desired level.

To summarize, the long-run discounting rate can be much lower than the mean, and indeed can correspond to low interest rates that are rarely observed. This dramatically illustrates the folly of assuming that the average real interest rate is the correct long-run discount rate. 

\begin{figure*}
\vspace*{.05in}
\centering
\includegraphics[width=14cm]{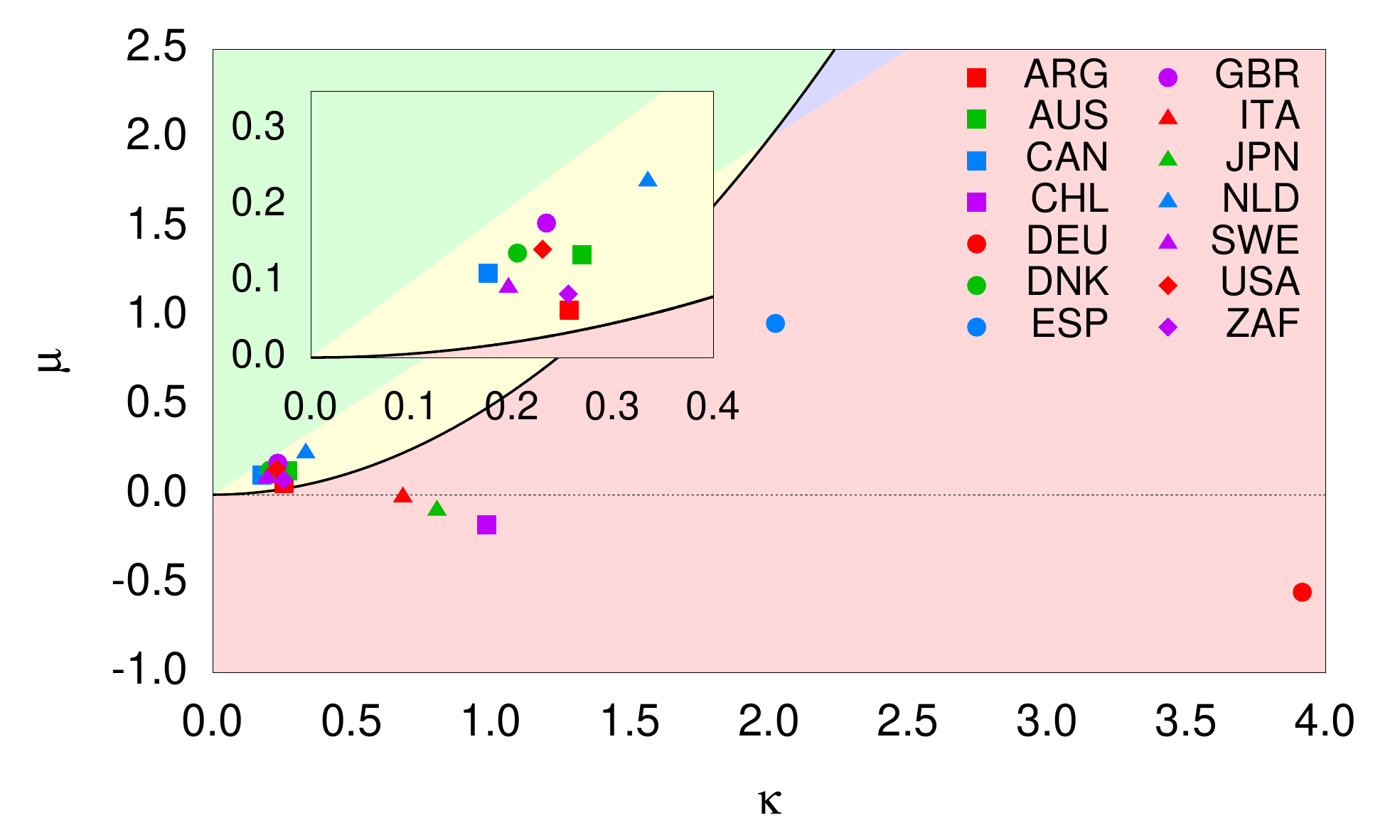}
\caption{A comparison of the parameters of the Ornstein-Uhlenbeck real interest rate model for the fourteen countries in our sample. The vertical axis is the non-dimensional mean interest rate $\mu=m/\alpha$ and the horizontal axis is the non-dimensional fluctuation amplitude $\kappa=k/\alpha^{3/2}$. Points to the upper left of the solid black curve have long-run discount rate $r_\infty > 0$, whereas for those in the lower right $r_\infty < 0$, i.e. the discount function $D(t)$ actually {\it increases} with time. While the discounting behavior of the nine stable countries is very similar, as shown in the inset, the other five countries behave very differently. Nonetheless, all fourteen countries are below the identity line (the green-yellow interface), indicating that in every case negative real interest rates are common.\label{phasePlane}}
\end{figure*}

Up to a rescaling of time, the long-run behavior of the model depends only on the two non-dimensional parameters $\mu$ and $\kappa$, defined as
\begin{equation}
\mu=\frac{m}{\alpha}, \qquad \kappa=\frac{k}{\alpha^{3/2}}.
\label{mu}
\end{equation}
The parameter space can be divided into two regions, as shown in Fig~\ref{phasePlane}. For the region in the upper left, where $\mu>\kappa^{2}/2$ (or equivalently $m>k^{2}/2\alpha^{2}$), the mean interest rate is large in comparison to the noise. The long-run discounting function decays exponentially at rate $r_\infty > 0$. 
For the region in the lower right $\mu<\kappa^{2}/2$ and thus $r_\infty < 0$, meaning the discount function $D(t)$ \textit{increases} exponentially. On the boundary, $m=k^{2}/2\alpha^{2}$, the long run interest rate $r_\infty =0$ and the discount function is asymptotically constant.

How is it possible for the discount function $D(t)$ to increase? This is easy to understand when there are persistent periods of negative real interest rates $r(t)$. Computation of the discount function $D(t)$ in Eq. (1) involves an average over exponentials, rather than the exponential of an average. As a result, periods where interest rates are negative are greatly amplified and can easily dominate periods where interest rates are large and positive, even if the negative rates are rarer and weaker (see Appendix). It does not take many such periods to produce long-run exponential growth of $D(t)$.

More surprising, Eq.~(\ref{r_inf}) shows that it is possible to get a negative long-run discounting rate even if negative real rates are rare. This occurs when $\mu>\kappa$, or equivalently, when $m > k/\alpha^{1/2}$, corresponding to the identity line in Fig~\ref{phasePlane}. On the other hand, from Eq.~(\ref{r_inf}), if $m < k^2/2\alpha^2$ then $r_\infty < 0$; if we keep the ratio $k/\alpha$ fixed while making $\alpha$ sufficiently small, then $r$ will rarely be negative (indeed, in this case $\mu/\kappa \propto \alpha^{-1/2} \gg 1$ as $\alpha \rightarrow 0$). The region where this is true corresponds to the blue wedge in the upper middle region of Fig~\ref{phasePlane}.


We fit the parameters of the OU model to each of the data series as described in the Appendix. The resulting parameters are listed in Table \ref{tab3}, and the position $(\kappa, \mu)$ of each country is shown in Fig~\ref{phasePlane}. The countries divide into two very clear groups. Nine countries, with relatively stable real interest rates, have long-run positive rates. They are in the exponentially decaying region to the upper left and are tightly clumped together near the zero
long-run interest rate curve. Five countries with less stable behavior, in contrast, are in the exponentially increasing region, which implies they have long-run negative rates, and are widely scattered. (It may not be a coincidence that all five have experienced fascist governments). In four cases the average log interest rate $m$ is negative due to at least one period of runaway inflation; the exception is Spain, which has a (highly positive) mean real interest rate, but still has a long-run negative rate. Note that all fourteen countries are below the identity line in Fig~\ref{phasePlane}, indicating that negative real interest rates are common -- even in the stable countries they occur $23\%$ of the time.
 
\begin{figure}
\vspace*{.05in}
\centering
\includegraphics[width=8.9cm]{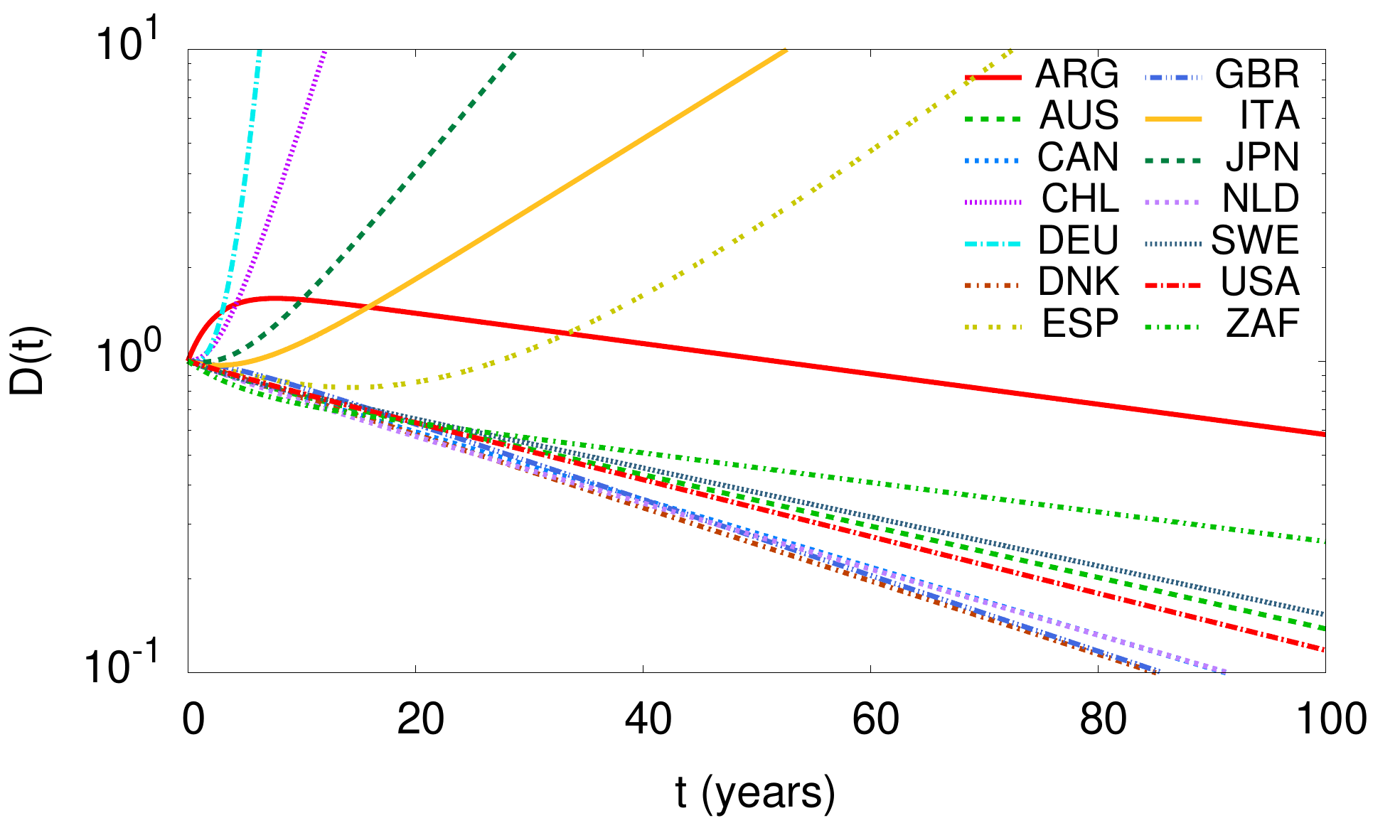}
\caption{The discounting function $D(t)$ as a function of time for the Ornstein-Uhlenbeck model for the fourteen countries in our sample. $D(t)$ quickly reaches its long-run exponential behavior. The long-run rates of the unstable countries vary dramatically, while most of the stable countries are fairly similar.\label{discountExamples}}
\end{figure}

In Fig.~\ref{discountExamples} we show the discount function $D(t)$
for all countries as a function of time, illustrating the
dramatic difference between the two groups. In most cases the behavior is monotonic; however, it can also be non-monotonic, as illustrated by Argentina, which initially increases and then decreases. 

In every case convergence to the long-run rate happens within 30 years, and typically within less than a decade. This is in contrast to other treatments of fluctuating rates, which assume short term rates are always (or nearly always) positive and predict that the decrease in the discounting rate happens over a much longer timescale, which can be measured in hundreds or thousands of years \cite{Weitzman98,NewellPizer,Gollier,Groom,Farmer,Hepburn,Freeman}.

To provide an estimate of statistical fluctuations we break each country's data into four equal sized blocks and estimate the parameters for each block separately. We quote the maximum and minimum values for each country in Table \ref{tab3}. This analysis reveals that statistical uncertainty is large. Focusing on the long-run interest rate
$r_\infty$, all countries have positive maximums and most have negative minimums -- only the USA, UK, and Denmark have positive $r_\infty$ in all four samples. Sub-sample variations are more than an order of magnitude larger than standard errors, indicating strong non-stationarity. 

\section{Discussion}

Our analysis here demonstrates that since real interest rates are often negative, and indeed are observed more than a quarter of the time, one must use a model that is compatible with this property. For this purpose we use the Ornstein-Uhlenbeck model, which we solve analytically. Our solution makes it easy to understand why the long-run discount rate is so low. The first reason is that real interest rates are typically fairly low. The average over all countries is $0.71\%$, and even the average over stable countries (those with $r_\infty > 0$) is $2.7\%$. The second reason is that the fluctuation term in the second part of Eq.\ (\ref{r_inf}), which depends both on the fluctuation amplitude $k$ and the persistence term $1/\alpha$, typically lowers rates for the stable countries by about $22\%$. In some cases, such as Spain, the effect is much more dramatic: Even though the mean short term rate has the high value of $m = 5.7\%$, the long-term discounting rate is $r_\infty = -6.4\%$. Averaging over the five unstable countries the mean interest rate $m = -2.9\%$ but $r_\infty = -42\%$. 

Our analysis here makes several simplifications, such as ignoring non-stationarity and correlations between the environment and the economy. We believe that including these effects, as we hope to do in future work, will only drive the discounting rate closer to zero. The methods that we have introduced here provide a foundation on which to incorporate more realistic assumptions.

We do not mean to imply that it is realistic to actually use the
increasing discounting functions that occur for the five countries with less stable interest rate processes. 
There is some validity to treating hyper-inflation as an aberration -- when it occurs government bonds are widely abandoned in favor of more stable carriers of wealth such as land and gold, and as a result under such circumstances the difference between nominal interest and inflation may underestimate the actual real rate of interest.

Nonetheless, the real interest rate is typically closely related to economic growth, and economic downturns are a reality. The great depression lasted for 15 years, and the fall of Rome triggered a depression in western Europe that lasted almost a thousand years. In light of our results here, arguments that we should wait to act on global warming because future
economic growth will easily solve the problem should be viewed with extreme skepticism. Our analysis clearly supports Stern over Nordhaus: Even if we throw out the five countries where we found asymptotically increasing discounting functions, the average long-run interest rate 
$r_\infty$ for the remaining nine countries is about $2\%$, only slightly more than the $1.4\%$ used by Stern. When we plan for the future we should always bear in mind that sustained economic downturns may visit us again, as they have in the past.

\appendix

\section{The discount function}

From Eq. (\ref{D}) the discount function can be written
$$
D(t)=E\left[e^{-x(t)}\right],
$$
where $x(t)$ is the random process
$$
x(t)=\int_0^t r(t')dt'
$$
representing the cumulative return at time $t$. Therefore,
\begin{equation}
D(t)=\int_{-\infty}^{\infty}dr\int_{-\infty}^{\infty} e^{-x}p(x,r,t|r_0)dx,
\label{D_2}
\end{equation}
where $p(x,r,t|r_0)$ is the joint probability density function of the bidimensional diffusion process $(x(t),r(t))$. Since $dx(t)=r(t)dt$ we see from Eq. (\ref{dr}) that the joint density obeys the following Fokker-Planck equation 
\begin{equation}
\frac{\partial p}{\partial t}=-r\frac{\partial p}{\partial x}+\alpha\frac{\partial}{\partial r}[(r-m)p]+\frac{k^2}{2}\frac{\partial^2 p}{\partial r^2},
\label{fpe}
\end{equation}
with the initial condition 
\begin{equation}
p(x,r,0|r_0)=\delta(x)\delta(r-r_0).
\label{initial_0}
\end{equation}
The problem is more conveniently addressed by working with the characteristic function, that is, the Fourier transform of the joint density 
\begin{eqnarray}
\tilde p(\omega_1,\omega_2,t|r_0)&=&\int_{-\infty}^\infty e^{-i\omega_1x}dx \label{cf}\\
&&\times\int_{-\infty}^\infty e^{-i\omega_2r}p(x,r,t|r_0)dr. \nonumber
\end{eqnarray}
Transforming Eqs. (\ref{fpe})-(\ref{initial_0}) results in the simpler equation:
$$
\frac{\partial\tilde p}{\partial t}=(\omega_1-\alpha\omega_2)\frac{\partial\tilde p}{\partial\omega_2}-\left(im\omega_2+\frac{k^2}{2}\omega_2^2\right)\tilde p,
$$
with
$$
\tilde p(\omega_1,\omega_2,0|r_0)=e^{-i\omega_2 r_0}.
$$
The solution of this initial-value problem is given by the Gaussian function
\begin{eqnarray}
\tilde p(\omega_1,\omega_2,t)&=&\exp\Bigl\{-A(\omega_1,t)\omega_2^2\nonumber \\
&-& B(\omega_1,t)\omega_2-C(\omega_1,t)\Bigr\},
\label{gaussian}
\end{eqnarray}
where the expressions for $A(\omega_1,t)$, $B(\omega_1,t)$, and $C(\omega_1,t)$ will be presented elsewhere. 

Once we have the characteristic function $\tilde p$ obtaining the equivalent discount function is straightforward. In effect, from Eqs. (\ref{D_2}) and (\ref{cf}) we see that
$$
D(t)=\tilde p\bigl(\omega_1=-i, \omega_2=0,t\bigr).
$$
In our case $D(t)=\exp\{-C(-i,t)\}$ which, after using the expression for $C(\omega_1,t)$ (to be detailed elsewhere) finally results in 
\begin{eqnarray}
\ln D(t)&=&-\frac{r_0}{\alpha}\left(1-e^{-\alpha t}\right)+\frac{\kappa^2}{2}\biggl[\alpha t \nonumber\\ 
&-&2\left(1-e^{-\alpha t}\right)
+\frac 12\left(1-e^{-2\alpha t}\right)\biggr] \nonumber\\
&-&\mu\left[\alpha t-\left(1-e^{-\alpha t}\right)\right].
\label{D_OU}
\end{eqnarray}
The exponential terms in Eq. (\ref{D_OU}) are negligible for large times ($t \gg \alpha^{-1}$). Finally, as $t \to \infty$, we get
\begin{equation}
\ln D(t)\simeq -(\mu-\kappa^2/2)\alpha t,
\label{D_OU_2}
\end{equation}
which is Eq. (\ref{assymptotic_D}). 

\begin{figure}[h]
\vspace*{.05in}
\centering
\includegraphics[width=8.9cm]{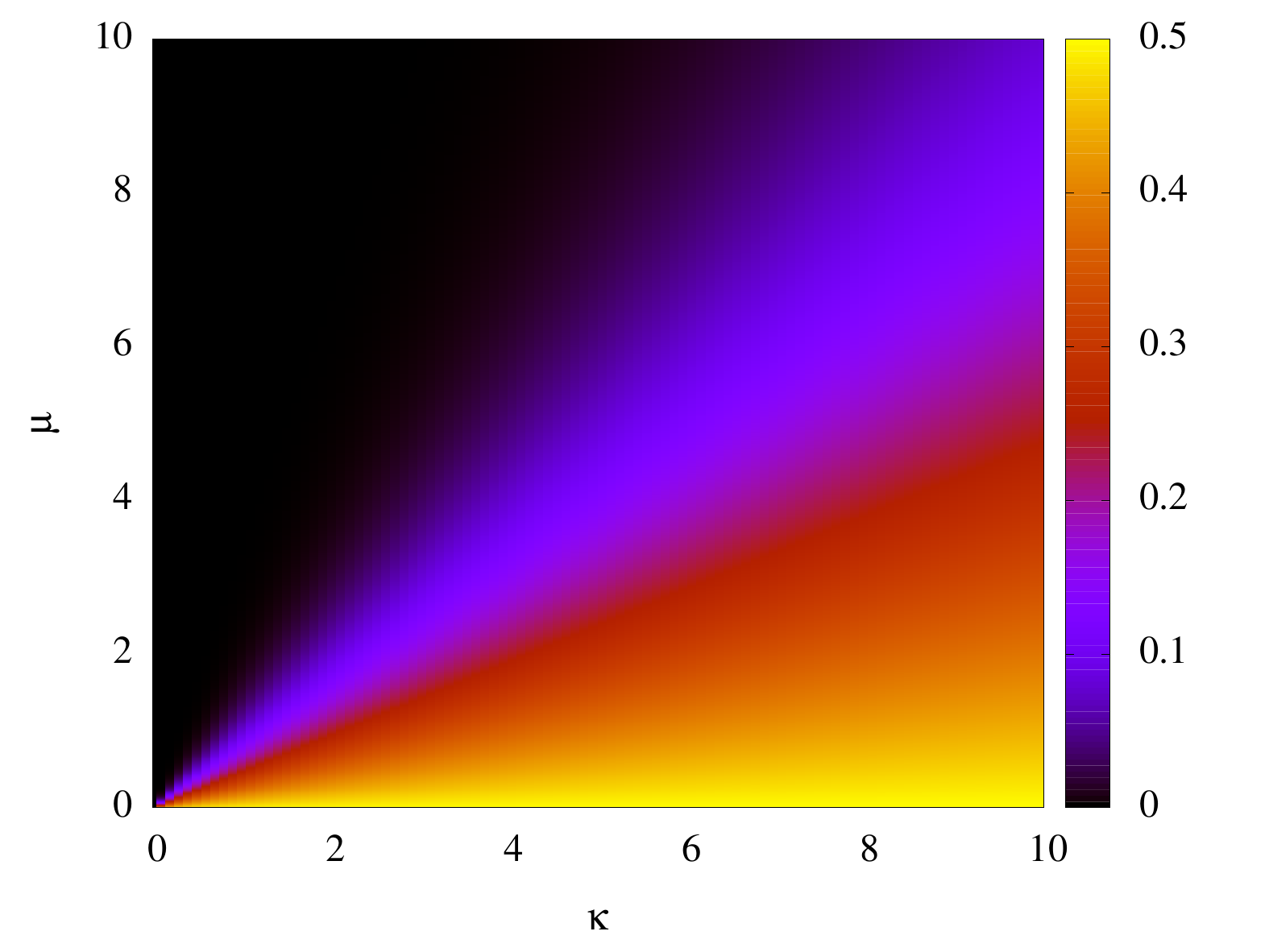}
\caption{The probability of negative rates as given in Eq. (\ref{p-stat2}). In the vicinity of the bottom right corner the probability of negative rates is around $0.5$ while at the upper left corner this probability is exponentially small and rates are mostly positive.\label{pminus}}
\end{figure}

\section{Negative rates}

As we have mentioned above the OU model may attain negative rates. Let us now quantify this characteristic by evaluating the probability $P(r<0,t|r_0)$, for $r(t)$ to be negative. It can be easily shown that the stationary probability, defined as
$$
 P_s^{(-)}=\lim_{t\rightarrow\infty}P(r<0,t|r_0),
$$ 
is given by
\begin{equation}
P_s^{(-)}=\frac 12{\rm Erfc}\left(\mu/\kappa\right),
\label{p-stat2}
\end{equation}
where ${\rm Erfc}(x)$ is the complementary error function and $\mu$ and $\kappa$ are defined in Eq. (\ref{mu}). Using standard asymptotic expressions of ${\rm Erfc}(x)$ we can easily obtain the behavior of $P_s^{(-)}$ in the cases (i) $\mu<\kappa$ and (ii) $\mu>\kappa$. Thus, (i) if the non-dimensional mean interest rate $\mu$ is smaller than the non-dimensional fluctuation amplitude $\kappa$, $\mu/\kappa<1$ and we have
\begin{equation}
P_s^{(-)}=\frac 12-\frac{1}{\sqrt\pi}(\mu/\kappa)+O(\mu^2/\kappa^2).
\label{p-i}
\end{equation}
For $\mu/\kappa$ sufficiently small, this probability approaches $1/2$. In other words, rates are positive or negative with almost equal probability. Note that this corresponds to the situation in which noise dominates over the mean value (see the low right corner in Fig. \ref{pminus}). 

(ii) When fluctuations around the normal level are smaller than the normal level itself, $\kappa<\mu$, we get
\begin{equation}
P_s^{(-)}\sim\frac{1}{2\sqrt\pi}\left(\frac{\kappa}{\mu}\right)e^{-\mu^2/\kappa^2}.
\label{p-ii}
\end{equation}
Therefore for mild fluctuations around the mean the probability of negative rates is {\it exponentially small} (see upper left corner of Fig. \ref{pminus}). 

In the limiting case where noise is balanced by the mean value, the probability of negative rates is $P_s^{(-)}=0.079$ and, due to the ergodic character of the OU process, this means that when $\mu=\kappa$ rates spend, on average, $7.9\ \%$ of the time with negative values.

\section{Rates below the long-run rate}

The probability that real rates $r(t)$ are below the long-run rate $r_\infty$ is given by
$$
P_\infty(t)\equiv{\rm Prob}\{r(t)<r_\infty\}=\int_{-\infty}^{r_\infty} p(r,t|r_0).
$$
It can be easily shown that in the stationary regime, $t \to \infty$, this probability is given by
\begin{equation}
P_\infty=\frac 12 {\rm Erfc}\left(\sqrt{\frac{m-r_\infty}{2\alpha}}\right).
\label{P_inf}
\end{equation}
Note that this expression proves that $P_\infty$ can be made arbitrarily small by increasing the persistence parameter $1/\alpha$ while holding $m$ and $r_\infty$ fixed. Indeed, using the asymptotic estimate 
$$
{\rm Erfc} (x)\sim \frac{e^{-x^2}}{\sqrt\pi x}\left[1+O\left(\frac{1}{x^2}\right)\right],
$$
we have
$$
P_\infty\sim\sqrt{\frac{\alpha}{2\pi(m-r_\infty)}}e^{-(m-r_\infty)/2\alpha},
$$
which is exponentially small when $\alpha \to 0$.

\section{Parameter estimation}

Real rates are nominal rates corrected by inflation. Nominal rates are given by the IG rates (i.e., 10 year Government Bond Yield) except in the cases of Chile and United Kingdom where, due to unavailability, we take the ID rates (i.e., the 10 year Discount rate). We transform the open IG or ID annual rates into logarithmic rates and denote the resulting time series by $b(t)$. Inflation is represented by the Consumer Price Index (CPI) and its log-rate is
$$
c(t)=\frac 1T\ln\left[C(t+T)/C(t)\right],
$$
where $T=10$ years and $C(t)$ is the time series of the empirical CPI for each country. Finally, the real interest rate, $r(t)$, is defined by
$$
r(t)=b(t)-c(t).
$$
The recording frequency for each country is either annual or quarterly. 

We estimate the parameters $m$, $k$ and $\alpha$ of the OU model as follows: The rate $m$ is the stationary average of the process (\ref{dr}):
$$
{\rm E}[r(t)]=m.
$$ 
We estimate $\alpha$ and $k$ based on the autocorrelation function $K(t-t')= {\rm E}\left[(r(t)-m)(r(t')-m)\right].$ For the OU process this is 
$$
K(t-t')=\frac{k^2}{2\alpha} e^{-\alpha|t-t'|},
$$
and $\alpha^{-1}$ is the correlation time. We estimate $\alpha$ (measured in units of 1/year) by evaluating the empirical auto-correlation and fitting it with an exponential. Once $\alpha$ is determined the parameter $k$ is obtained from the (empirical) standard deviation, $\sigma^2={\rm E}\left[(r(t)-m)^2\right],$
which is given by the correlation function since $\sigma^2=K(0)$. Hence 
$$
k=\sigma\sqrt{2\alpha}.
$$

In order to have an idea about the robustness of the estimation procedure we split the constructed real interest rate data from each country into four equally spaced blocks. In each block we estimate the parameters of the OU model applying the method described above, except for the parameter $\alpha$, which is always estimated using the complete data set. The main reason to avoid estimating $\alpha$ on small blocks is because the time series of some countries are too short. Instead the quoted uncertainty in $\alpha$ is the standard least square error, computed by fitting an exponential to the autocorrelation function of the real interest time series. Table \ref{tab3} shows the minimum and the maximum values for $\mu$, $\kappa$ and $r_\infty$, and their uncertainties under subsampling. 

\acknowledgments

We would like to thank National Science Foundation grant 0624351. We also acknowledge partial support form the Ministerio de Ciencia e Innovaci\'on under contract No. FIS2009-09689 and the Institute for New Economic Thinking.


\begin{thebibliography}{99}
\bibitem {Farmer} Farmer JD, Geanakoplos J, Hyperbolic discounting is rational: Valuing the far future with uncertain discount rates, SSRN.
\end{thebibliography}
\end{document}